\documentclass[10pt,amsmath,amssymb,aps,pra,twocolumn,showpacs,showkeys,groupedaddress]{revtex4-1}

\pdfoutput=1

\usepackage{graphicx}  
\usepackage{dcolumn}   
\usepackage{bm}        

\usepackage[colorlinks=true,urlcolor=blue,bookmarks=false]{hyperref}  

\usepackage{color}

\usepackage{verbatim}
\usepackage{amsmath}
\usepackage{latexsym}
\usepackage{amsfonts}
\usepackage{amssymb}

\begin{document}

\title{Measurement of Anomalous Moments of Creation and Annihilation Operators}

\author{Ranjana Prakash}
\email{prakash\_ranjana1974@rediffmail.com}
\author{Ajay K. Yadav}
\email{ajaypdau@gmail.com}
\affiliation{Physics Department, University of Allahabad, Allahabad-211002, UP, India.}


\begin{abstract}
We propose a scheme for measurements of anomalous moments of creation and annihilation operators requiring only one beam splitter, only one photodetector and not many measurements with phase-shift $\varphi$ in homodyning. This has advantageous over scheme of Shchukin and Vogel [Phys. Rev. A {\bf 72}, 043808 (2005)] which requires many photodetectors, many beam splitters and a large number of measurements with phase-shift $\varphi$ as a Fourier transform is calculated integrating over $\varphi$.
\end{abstract}



\maketitle

\section{Introduction}
Classical theory is inadequate for explanation of properties of states, for which the weight function in diagonal representation \cite{pfunction} is not non-negative or becomes more singular than a delta function and hence cannot interpreted as a probability distribution. Antibunching \cite{antibunch}, sub-Poissonian photon statistics \cite{david} and various kinds of squeezing \cite{squeeze} are some of the nonclassical features of optical fields. Earlier, study of such nonclassical effects was largely in academic interest \cite{noncleff}, but after the demonstration of photon antibunching \cite{kimbleetal}, sub-Poissonian photon statistics \cite{sm} and squeezing \cite{slusheretal}, their applications in quantum information processing such as quantum teleportation \cite{qt}, quantum dense coding \cite{qdc}, quantum cryptography \cite{hillery1}, their importance is now well realized.

For the measurements of antibunching \cite{kimbleetal} and sub-Poisson photon statistics \cite{sm}, photon counting techniques can be used which gives direct measurement of the intensity fluctuations of a light field. Measurement of squeezing is done by homodyning of squeezed light with coherent light whose phase can be varied and by measuring the fluctuations in superposed light with one or more photodetectors, e.g., ordinary homodyne detection \cite{mandel}, homodyne cross correlation \cite{ouetal}, homodyne intensity correlation \cite{vogel}, etc. In study of squeezing in the resonance fluorescence from a single trapped and cooled ion based on the observation of the photon pair correlations by homodyne detection anomalous moments \cite{vogel, vogel1} were observed using weak local oscillator. Here "weak" means intensity of the local oscillator is of the same order as magnitude of the fluorescence intensity, and anomalous moments are expectation values containing different numbers of annihilation and creation operators. These anomalous moments of the fluorescence also contribute to the non-classical behavior of the light in homodyne detection. Prakash and Kumar \cite{pk} proposed a balanced homodyne method for detection of fourth-order squeezing. Prakash and Mishra \cite{pm} extended the proposal of ordinary homodyning for experimental detection of amplitude-squared squeezing by measuring higher order moments of number operator of mixed light with shifted phases. They also studied higher-order sub-Poissonian photon statistics conditions for non-classicality and discussed its use for the detection of Hong and Mandel's squeezing of arbitrary order. Prakash, Kumar and Mishra reported \cite{pkm} an ordinary homodyne method for detection of second type of amplitude-squared squeezing of Hillery. Prakash and Yadav reported recently \cite{py} an ordinary homodyne method for detection of amplitude nth-power squeezing of by measuring the higher-order factorial moments of the number operator in light obtained by homodyning with coherent light with shifted phases.

Since the diagonal coherent-state representation is not practically accessible, the nonclassicality criterion must be reformulated in terms of observable quantities, e.g., normally-ordered moments. Agarwal and Tara \cite{at} described a quantitative criterion involving normally ordered moments for characterizing the nonclassical properties for states which may not exhibit squeezing or sub-Poissonian statistics. Agarwal \cite{agarwal} also described nonclassical characteristics of marginals. Conditions for the nonclassicality of quantum states in terms of moments of the creation and annihilation operators, of two quadratures, and of a quadrature and a photon number operator have been formulated \cite{srv, sv} and necessary and sufficient conditions for nonclassicality were obtained. Authors showed that all required moments could be determined by homodyne correlation measurements with weak local oscillator. These methods of characterizing non-classical effects by moments of annihilation and creation operators have been applied to the characterization of amplitude-squared squeezing \cite{sv}. Shchukin and Vogel \cite{sv1} have also proposed a method based on balanced homodyne correlation measurement for measuring general space-time-dependent correlation functions of quantized radiation fields. Also, they showed that the condition for entanglement is based on moments involving unequal powers of photon annihilation and creation operators \cite{vs}. But in their proposed methods of measuring anomalous moments, number of required beam splitters and photo-detectors increased with the increase in the order of moments. Also, a very large number of repeated measurements with phase shift $\varphi$ of the local oscillator are necessary as evaluation of Fourier transform of a function of $\varphi$ is involved. This motivates us to give a simpler scheme for the measurement of anomalous moments of creation and annihilation operators.

We present here a proposal, based on the observation of higher order factorial moments of photon number operator \cite{py}, for experimental detection of anomalous moments of creation and annihilation operators by using the ordinary homodyne detection method. Our detection method requires only one photodetector and only one beam splitter, and it works without imposing any condition on coherent state and transmittance/reflectance of the beam splitter.

\section{The detection scheme}
\begin{figure}
	\centering
		\includegraphics[width=0.48\textwidth]{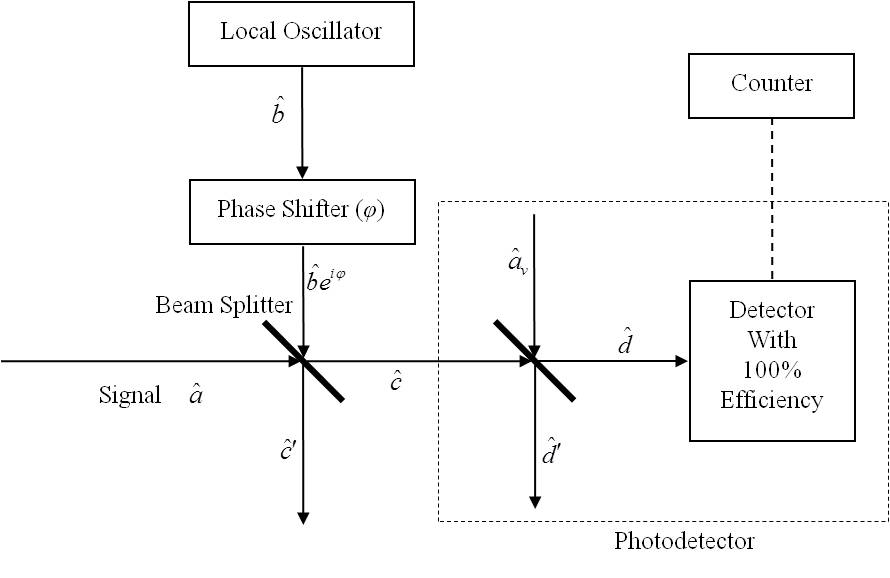}
	\caption{\label{dfmoh}Schematic diagram of detection of moments via ordinary homodyning.}
\end{figure}

We propose the same detection scheme which we proposed earlier \cite{py}  for study of amplitude $n$th-power squeezing. It is shown in Fig.~\ref{dfmoh}, and it brings out \cite{lk} the conceptual meaning of quantum efficiency of the experimental detector. A beam splitter and an ideal detector placed inside the dotted rectangle model a real photodetector. A beam splitter mixes the signal represented by operator $\hat{a}$ with signal $\hat{b}e^{i\varphi}$ obtained by shifting by $\varphi$ the phase of signal from a local oscillator represented by operator $\hat{b}$ to give output signals $\hat{c}$ and $\hat{c}'$. If the beam splitter has transmittance $T$ and we write as $t = \sqrt{T}$ and $r = \sqrt{1-T}$ coefficients of transmission and reflection for the amplitudes, respectively, we can write one output signal \cite{lp} as $\hat{c} = t \hat{a} + r \hat{b} e^{i\varphi},$ with $t$ and $r$ real. Number operator of the mixed light is then given by $\hat{N}_{c} = \hat{c}^{\dagger} \hat{c} = (t \hat{a}^{\dagger} + r \hat{b}^{\dagger} e^{-i\varphi})(t \hat{a} + r \hat{b} e^{i\varphi})$.

If a beam splitter mixes input signal $\hat{c}$ and a vacuum signal $\hat{a}_{v}$ (see Fig.~\ref{dfmoh}), one of the outputs $\hat{d}$ given by $\hat{d} = \sqrt{\eta} \hat{c} + \sqrt{1-\eta} \hat{a}_{v}$ detected by an ideal detector having $100\%$ efficiency models \cite{lk} a realistic detector with quantum efficiency $\eta$. In this model the detected counts, $\langle \hat{d}^{\dagger} \hat{d} \rangle$, are $\eta$ times the incident number of photons, $\langle \hat{c}^{\dagger} \hat{c} \rangle$, i.e., $\langle \hat{d}^{\dagger} \hat{d} \rangle = \eta \langle \hat{c}^{\dagger} \hat{c} \rangle$ and factorial moments of order $n$, $\langle \hat{d}^{\dagger n} \hat{d}^{n} \rangle = \eta^{n} \langle \hat{c}^{\dagger n} \hat{c}^{n} \rangle$.

For the setup under consideration, the observed factorial moments of counts with phase shift $\varphi$ is then
\begin{eqnarray}
M^{(n)}_{\varphi} &=& \eta^{n} \langle\hat{c}^{\dagger n} \hat{c}^{n}\rangle \nonumber \\
&=& \eta^{n} \sum^{n}_{l,m=0} {^{n}C_{l}} {^{n}C_{m}} t^{2n-l-m} r^{l+m} \nonumber \\
&& \langle \hat{a}^{\dagger n-l} \hat{a}^{n-m} \hat{b}^{\dagger l} \hat{b}^{m} \rangle e^{i(m-l)\varphi}.
\end{eqnarray}
If observations are done for $M^{(n)}_{\varphi} e^{\pm id\varphi}$ for $\varphi = k\pi/n$ with $k = 0, 1,..., 2n-1$ and for $d$ such that $0 \leq d \leq n-1$, we can easily find the values of observables
\begin{eqnarray}
R_{n}(d) &=& (2n\eta^{n}t^{2n}\xi^{d})^{-1}e^{id\theta_{\beta}}\sum^{2n-1}_{k=0}M^{(n)}_{k\pi/n}e^{idk\pi/n} \nonumber \\
&=& \sum^{n-d}_{l=0}{^{n}C_{l}}{^{n}C_{d+l}}\xi^{2l}\langle \hat{a}^{\dagger n-d-l}\hat{a}^{n-l} \rangle, \\
\label{rn}
R'_{n}(d) &=& [R_{n}(d)]^{*} \nonumber \\
&=& \sum^{n-d}_{l=0}{^{n}C_{l}}{^{n}C_{d+l}}\xi^{2l}\langle \hat{a}^{\dagger n-l}\hat{a}^{n-d-l} \rangle,
\label{rnp}
\end{eqnarray}
where $\beta \equiv |\beta| e^{i\theta_{\beta}}$ is the complex amplitude in coherent state $|\beta\rangle$ generated by the local oscillator, and $\xi \equiv t^{-1} r |\beta|$. For $d = n$ observations are done for $M^{(n)}_{\varphi} e^{\pm in\varphi}$ for $\varphi = k \pi /2n$ with $k = 0, 1,..., 4n-1$ and we can easily find the values of observables
\begin{eqnarray}
S_{n} &=& \frac{e^{in\theta_{\beta}}}{4n}\sum^{4n-1}_{k=0}M^{(n)}_{k\pi/2n}e^{ik\pi/2}=(\eta tr|\beta|)^{n}\langle \hat{a}^{n} \rangle, \label{sn}\\
S'_{n} &=& [S_{n}]^{*}=(\eta tr|\beta|)^{n}\langle \hat{a}^{\dagger n} \rangle.
\label{snp}
\end{eqnarray}

\section{Measurement of anomalous moments}
Clearly, moments $\langle \hat{a}^{n} \rangle$ and $\langle \hat{a}^{\dagger n} \rangle$ are obtained directly from Eqs.~(\ref{sn}) and ~(\ref{snp}), respectively. To achieve moments $\langle \hat{a}^{\dagger n-d} \hat{a}^{n} \rangle$, we can solve Eq.~(\ref{rn}) by iteration. This is done in Appendix and gives
\begin{equation}
{^{n}C_{d}}\langle \hat{a}^{\dagger n-d} \hat{a}^{n} \rangle = \sum^{n-d}_{s=0} K_{s} (^{n}C_{s})^{2} \xi^{2s} R_{n-s}(d),
\label{amrn}
\end{equation}
where $K_{s}$ is defined by
\begin{equation}
{^{d+s}C_{s}}K_{s}=-\sum^{s-1}_{l=0}K_{l}{^{s}C_{l}}{^{d+s}C_{l}} \text{ with } K_{0}=1.
\label{ks}
\end{equation}
Adopting similar method, as shown in Appendix, we can also solve Eq.~(\ref{rnp}) to achieve moments $\langle \hat{a}^{\dagger n} \hat{a}^{n-d} \rangle$. Also, this is given directly as
\begin{eqnarray}
{^{n}C_{d}}\langle \hat{a}^{\dagger n}\hat{a}^{n-d} \rangle &=& [{^{n}C_{d}}\langle \hat{a}^{\dagger n-d}\hat{a}^{n} \rangle]^{*} \nonumber \\
&=& \sum^{n-d}_{s=0}K_{s}(^{n}C_{s})^{2}\xi^{2s}R'_{n-s}(d),
\label{amrnp}
\end{eqnarray}
where $K_{s}$ is defined by Eq.~(\ref{ks}). Clearly, when $d = 0$ \cite{py}, Eqs.~(\ref{amrn}) and ~(\ref{amrnp}) are same and give
\begin{equation}
\langle \hat{a}^{\dagger n}\hat{a}^{n} \rangle=\sum^{n}_{s=0}K_{s}(^{n}C_{s})^{2}\xi^{2s}R_{n-s},
\end{equation}
where $K_{s}$ is obtained from Eq.~(\ref{ks}) by putting $d = 0$. For $d > 0$, we can obtain moments of unequal powers of creation and annihilation operators from Eqs.~(\ref{amrn}) and ~(\ref{amrnp}).

If $\hat{q}$ and $\hat{p}$ are quadrature operators defined by $\hat{q} = (\hat{a}^{\dagger} + \hat{a})/\sqrt{2}$, $\hat{p} = i(\hat{a}^{\dagger} - \hat{a})/\sqrt{2}$ and $\hat{n} = \hat{a}^{\dagger} \hat{a}$ is the number operator, normally ordered moments of the $l$th power of these operators can be written as
\begin{eqnarray}
\langle :\hat{q}^{l}:\rangle &=& \frac{1}{2^{l/2}} \sum^{l}_{m=0} {^{l}C_{m}} \langle \hat{a}^{\dagger l-m} \hat{a}^{m}\rangle, \\
\label{mq}
\langle :\hat{p}^{l}:\rangle &=& \frac{i^{l}}{2^{l/2}} \sum^{l}_{m=0} (-1)^{m} {^{l}C_{m}} \langle \hat{a}^{\dagger l-m} \hat{a}^{m}\rangle, \\
\label{mp}
\langle :\hat{n}^{l}: \rangle &=& \langle \hat{a}^{\dagger l} \hat{a}^{l} \rangle = \langle \hat{n}^{(l)} \rangle.
\label{mn}
\end{eqnarray}
Similarly, moment of normal ordering of operators $\hat{p}^{k} \hat{q}^{l}$, $\hat{q}^{k} \hat{n}^{l}$ and $\hat{p}^{k} \hat{n}^{l}$ are given by
\begin{eqnarray}
\langle :\hat{p}^{k} \hat{q}^{l}: \rangle &=& \frac{i^{k}}{2^{(k+l)/2}}\sum^{k}_{m_{1}=0}(-1)^{m_{1}}{^{k}C_{m_{1}}} \nonumber \\
&& \sum^{l}_{m_{2}=0}{^{l}C_{m_{2}}} \langle \hat{a}^{\dagger k+l-m_{1}-m_{2}}\hat{a}^{m_{1}+m_{2}}\rangle, \\
\label{ampq}
\langle :\hat{q}^{k} \hat{n}^{l}: \rangle &=& \frac{1}{2^{k/2}}\sum^{k}_{m=0}{^{k}C_{m}} \langle \hat{a}^{\dagger k+l-m}\hat{a}^{l+m}\rangle, \\
\label{amqn}
\langle :\hat{p}^{k} \hat{n}^{l}: \rangle &=& \frac{i^{k}}{2^{k/2}}\sum^{k}_{m=0} (-1)^{m} {^{k}C_{m}} \langle \hat{a}^{\dagger k+l-m}\hat{a}^{l+m}\rangle,
\label{ampn}
\end{eqnarray}
It is clear from Eqs.~(\ref{mq})-~(\ref{ampn}) that normally ordered such moments can be written in terms of moments of unequal powers of creation and annihilation operators and their values can be obtained by measuring anomalous moments of annihilation and creation operators.

\section{Discussion of results}
Quantum efficiency $\eta$ of the detector may be found using spontaneous parametric down conversion \cite{mqe}. If a photon of $\hbar \omega$ breaks to create two photons of energies $\hbar \omega_{1}$ and $\hbar \omega_{2}$ ($\omega_{1}+\omega_{2} = \omega$), the latter two photons give coincidence counts ideally. If $N$ photons break and the quantum efficiencies for modes $\omega_{1}$ and $\omega_{2}$ are $\eta_{1}$ and $\eta_{2}$ (both $<1$), the experiment should register counts $N_{1} = \eta_{1} N, N_{2} = \eta_{2} N$ and coincidence counts $N_{c} = \eta_{1} \eta_{2} N$. This gives $\eta_{1} = N_{c}/N_{2}$ and $\eta_{2} = N_{c}/N_{1}$.

In the scheme of Shchukin and Vogel, the authors arranged several beam splitters in several levels. In addition to a beam splitter at entrance of beam from local oscillator and one used for mixing with signal there are $2^{r}$ beam splitters at $r$th level. Hence for depth (maximum level) $d$ the number of required beam splitters is $2^{d}+1$. Also, for depth $d$, $2d$ photodetectors are required. For depth $d$, if $k$ satisfies $2^{n} \geq k > 2^{n-1}$, this scheme allows one to measure the moments for $k, l = 0, 1,..., 2^{d}$, step by step. For obtaining values of $\langle \hat{a}^{k} \rangle$ and $\langle \hat{a}^{\dagger k}\hat{a}^{k} \rangle$, thus, depth $d = n$ is needed,  and it requires use of $2^{n}+1$ beam splitters and $2n$ photodetectors. In contrast, in the method proposed in the present paper, only one beam splitter and only one photodetector is required. The Shchukin-Vogel method proposes measurement of correlations with several phase shifts $\varphi$ of local oscillator. Since a Fourier transform of correlations over $\varphi$ is required for inferring the values of moments $\langle \hat{a}^{\dagger k} \hat{a}^{l} \rangle$, a very large number of repeatition of experiments with changed values of $\varphi$ should be required. It may be noted that in the method proposed in the present paper only ${l(2k-l+1)+4(k-l)}$ repetitions are required for moments $\langle \hat{a}^{\dagger k}\hat{a}^{l} \rangle$ for $k > l$.

Since normally ordered moments of product of powers of the two quadratures, and product of powers of a quadrature and the photon number operator, can be written in terms of moments of powers of creation and annihilation operators, their values can also be obtained by measuring moments of annihilation and creation operators.

\begin{acknowledgments}
We would like to thank Prof. H. Prakash and Prof. N. Chandra for their interest and critical comments. Also we would like to acknowledge Dr. Rakesh Kumar, Dr. Pankaj Kumar, Dr. Devendra K. Mishra, Dr. Namrata Shukla, Dr. Manoj K. Mishra, Mr. Ajay K. Maurya and Mr. Vikram Verma for their valuable and stimulating discussions. One of the authors (AKY) is grateful to the University Grants Commission(UGC), New Delhi, India for financial support.
\end{acknowledgments}

\appendix*\section{}

We can separate the $\langle \hat{a}^{\dagger n-d}\hat{a}^{n} \rangle$ term on right hand side of Eq.~(\ref{rn}) and write
\begin{eqnarray}
{^{n}C_{d}}\langle \hat{a}^{\dagger n-d}\hat{a}^{n} \rangle = R_{n}(d) &&- \sum^{n-d}_{l=1}{^{n}C_{l}}{^{n}C_{d+l}}\xi^{2l} \nonumber \\
&& \langle \hat{a}^{\dagger n-d-l}\hat{a}^{n-l} \rangle.
\label{A1}
\end{eqnarray}
In the first term in summation on the right hand side we substitute for ${^{n-1}C_{d}}\langle \hat{a}^{\dagger n-d-1}\hat{a}^{n-1} \rangle$ the expression obtained from Eq.~(\ref{A1}) and this gives
\begin{equation*}
{^{n}C_{d}}\langle \hat{a}^{\dagger n-d}\hat{a}^{n} \rangle = R_{n}(d)+K_{1}(^{n}C_{1})^{2}\xi^{2}[R_{n-1}(d)
\end{equation*}
\begin{eqnarray}
&&-\sum^{n-1-d}_{m=1}{^{n-1}C_{m}}{^{n-1}C_{d+m}}\xi^{2m}\langle \hat{a}^{\dagger n-1-d-m}\hat{a}^{n-1-m} \rangle] \nonumber \\
&&-\sum^{n-d}_{l=2}{^{n}C_{l}}{^{n}C_{d+l}}\xi^{2l}\langle \hat{a}^{\dagger n-d-l}\hat{a}^{n-l} \rangle
\end{eqnarray}
with ${^{d+1}C_{1}}K_{1}=-1$. This can be simplified and written as
\begin{eqnarray}
&&{^{n}C_{d}}\langle \hat{a}^{\dagger n-d}\hat{a}^{n} \rangle = R_{n}(d) + K_{1}(^{n}C_{1})^{2}\xi^{2}R_{n-1}(d) \nonumber \\
&&-\sum^{n-d}_{l=2}[1+K_{1}{^{l}C_{1}}{^{d+l}C_{1}}]{^{n}C_{l}}{^{n}C_{d+l}}\xi^{2l}\langle \hat{a}^{\dagger n-d-l}\hat{a}^{n-l} \rangle.  \nonumber \\
\end{eqnarray}
If we again substitute the first term in summation the expression obtained for ${^{n-2}C_{d}}\langle \hat{a}^{\dagger n-d-2}\hat{a}^{n-2} \rangle$ from Eq.~(\ref{A1}) and simplify, we get
\begin{eqnarray}
&&{^{n}C_{d}}\langle \hat{a}^{\dagger n-d}\hat{a}^{n} \rangle = R_{n}(d)+K_{1}(^{n}C_{1})^{2}\xi^{2}R_{n-1}(d) \nonumber \\
&&+ K_{2}(^{n}C_{2})^{2}\xi^{4}R_{n-2}(d) -\sum^{n-d}_{l=3}[1+K_{1}{^{l}C_{1}}{^{d+l}C_{1}} \nonumber \\
&&+ K_{2}{^{l}C_{2}}{^{d+l}C_{2}}]{^{n}C_{l}}{^{n}C_{d+l}}\xi^{2l}\langle \hat{a}^{\dagger n-d-l}\hat{a}^{n-l} \rangle
\end{eqnarray}
with ${^{d+2}C_{2}}K_{2}=-[1+K_{1}{^{2}C_{1}}{^{d+2}C_{1}}]$. If we go on doing similar exercises we get required Eq.~(\ref{amrn}), where $K_{s}$ is defined by Eq.~(\ref{ks}).


\begin{thebibliography}{99}

\bibitem{pfunction}
E.C.G. Sudarshan, Phys. Rev. Lett. {\bf 10}, 277 (1963); R.J. Glauber, Phys. Rev. {\bf 131}, 2766 (1963).

\bibitem{antibunch}
N. Chandra and H. Prakash, Phys. Rev. A {\bf 1}, 1696 (1970); H. Paul, Rev. Mod. Phys. {\bf 54}, 1061 (1982).

\bibitem{david}
L. Davidovich, Rev. Mod. Phys. {\bf 68}, 127 (1996).

\bibitem{squeeze}
D.F. Walls, Nature {\bf 306}, 141 (1983); C.K. Hong and L. Mandel, Phys. Rev. Lett. {\bf 54}, 323 (1985); M. Hillery, Opt. Commun. {\bf 62}, 135 (1987); Z.-M. Zhang, L. Xu, J.-L. Chai and F.-L. Li, Phys. Lett. A {\bf 150}, 27 (1990).

\bibitem{noncleff}
B.R. Mollow and R.J. Glauber, Phys. Rev. {\bf 160}, 1076 (1967); {\bf 160}, 1097 (1967); N. Chandra and H. Prakash, Lett. Nuovo. Cim. {\bf 4}, 1196 (1970); M.T. Raiford, Phys. Rev. A {\bf 2}, 1541 (1970); N. Chandra and H. Prakash, Indian J. Pure Appl. Phys. {\bf 9}, 409 (1971); {\bf 9}, 677 (1971); {\bf 9}, 688 (1971); {\bf 9}, 767 (1971); H. Prakash, N. Chandra and Vachaspati, Phys. Rev. A {\bf 9}, 2167 (1974); H. Prakash, N. Chandra and Vachaspati, Indian J. Pure Appl. Phys. {\bf 13}, 757 (1975); {\bf 13}, 763 (1975); {\bf 14}, 41 (1976); {\bf 14}, 48 (1976).

\bibitem{kimbleetal}
H.J. Kimble, M. Dagenais and L. Mandel, Phys. Rev. Lett. {\bf 39}, 691 (1977).

\bibitem{sm}
R. Short and L. Mandel, Phys. Rev. Lett.  {\bf 51}, 384 (1983).

\bibitem{slusheretal}
R.E. Slusher et al, Phys. Rev. Lett. {\bf 55}, 2409 (1985).

\bibitem{qt}
S.L. Braunstein and H.J. Kimble, Phys. Rev. Lett. {\bf 80}, 869 (1998); J. Zhang and K.C. Peng, Phys. Rev. A {\bf 62}, 064302 (2000); W.P. Bowen, P.K. Lam and T.C. Ralph, J. Mod. Opt. {\bf 50}, 801 (2003).

\bibitem{qdc}
M. Ban, J. Opt. B {\bf 1}, L9 (1999); S.L. Braunstein and H.J. Kimble, Phys. Rev. A {\bf 61}, 042302 (2000).

\bibitem{hillery1}
M. Hillery, Phys. Rev. A {\bf 61}, 022309 (2000).

\bibitem{mandel}
L. Mandel, Phys. Rev. Lett. {\bf 49}, 136 (1982).

\bibitem{ouetal}
Z.Y. Ou, C.K. Hong and L. Mandel, Phys. Rev. A {\bf 36}, 192 (1987).

\bibitem{vogel}
W. Vogel, Phys. Rev. Lett. {\bf 67}, 2450 (1991).

\bibitem{vogel1}
W. Vogel, Phys. Rev. A {\bf 51}, 4160 (1995).

\bibitem{pk}
H. Prakash and P. Kumar, J. Opt. B {\bf 7}, S786 (2005).

\bibitem{pm}
H. Prakash and D.K. Mishra, J. Phys. B {\bf 39}, 2291 (2006); {\bf 40}, 2531 (2007).

\bibitem{pkm}
H. Prakash, P. Kumar, D.K. Mishra, Int. J. Mod. Phys. B {\bf 24}, 5547 (2010).

\bibitem{py}
R. Prakash and A.K. Yadav, Opt. Commun. {\bf 285}, 2387 (2012).

\bibitem{at}
G.S. Agarwal and K. Tara, Phys. Rev. A {\bf 46}, 485 (1992).

\bibitem{agarwal}
G.S. Agarwal, Opt. Commun. {\bf 95}, 109 (1993).

\bibitem{srv}
E. Shchukin, Th. Richter and W. Vogel, Phys. Rev. A {\bf 71}, R011802 (2005).

\bibitem{sv}
E.V. Shchukin and W. Vogel, Phys. Rev. A {\bf 72}, 043808 (2005).

\bibitem{sv1}
E. Shchukin and W. Vogel, Phys. Rev. Lett. {\bf 96}, 200403 (2006).

\bibitem{vs}
W. Vogel and E. Shchukin, J. Phys.: Conf. Series {\bf 84}, 012020 (2007).

\bibitem{lk}
See, e.g., the review article R. Loudon and P.L. Knight, J. Mod. Opt. {\bf 34}, 709 (1987).

\bibitem{lp}
See, e.g., the review article U. Leonhardt and H. Paul, Prog. Qunat. Electr. {\bf 19}, 89 (1995).

\bibitem{mqe}
S. Castelletto, A. Godone, C. Novero and M.L. Rastello, Metrologia {\bf 32}, 501 (1995); S. Lu, B. Liu, B. Sun, Z. Xu and D. Jiang, Meas. Sci. Technol. {\bf 13}, 186 (2002).

\end{thebibliography}
\end{document}